\documentclass[
 aip,
  secnumarabic,
preprintnumbers,
  amssymb,
  amsmath]{revtex4-1}
\usepackage{graphicx}
\usepackage{dcolumn}
\usepackage{bm}
\usepackage{hyperref}
\usepackage[small]{titlesec}
\usepackage{ifthen}
\usepackage{color}
\usepackage{ulem}             
\usepackage{proof}

\newif\ifcom
\newif\ifdel
\comtrue               %
\titleformat{\section}{\normalfont\sffamily\bfseries\color{black}} {\thesection}{12pt}{}
\delfalse
\begin{document}

\title{Anomalous Hall effect in YIG$|$Pt bilayers} 

\author{Sibylle Meyer}
\email{sibylle.meyer@wmi.badw.de}
\affiliation{Walther-Mei{\ss}ner-Institut, Bayerische Akademie der Wissenschaften, 85748 Garching, Germany}
\affiliation{Physik-Department, Technische Universit\"{a}t M\"{u}nchen, 85748 Garching, Germany}
\author{Richard Schlitz}
\affiliation{Walther-Mei{\ss}ner-Institut, Bayerische Akademie der Wissenschaften, 85748 Garching, Germany}
\affiliation{Physik-Department, Technische Universit\"{a}t M\"{u}nchen, 85748 Garching, Germany}
\author{Stephan Gepr\"{a}gs}
\affiliation{Walther-Mei{\ss}ner-Institut, Bayerische Akademie der Wissenschaften, 85748 Garching, Germany}
\author{Matthias Opel}
\affiliation{Walther-Mei{\ss}ner-Institut, Bayerische Akademie der Wissenschaften, 85748 Garching, Germany}
\author{Hans Huebl}
\affiliation{Walther-Mei{\ss}ner-Institut, Bayerische Akademie der Wissenschaften, 85748 Garching, Germany}
\affiliation{Nanosystems Initiative Munich, 80799 M\"{u}nchen, Germany}
\author{Rudolf Gross}
\affiliation{Walther-Mei{\ss}ner-Institut, Bayerische Akademie der Wissenschaften, 85748 Garching, Germany}
\affiliation{Physik-Department, Technische Universit\"{a}t M\"{u}nchen, 85748 Garching, Germany}
\affiliation{Nanosystems Initiative Munich, 80799 M\"{u}nchen, Germany}
\author{Sebastian T. B. Goennenwein}
\affiliation{Walther-Mei{\ss}ner-Institut, Bayerische Akademie der Wissenschaften, 85748 Garching, Germany}
\affiliation{Nanosystems Initiative Munich, 80799 M\"{u}nchen, Germany}
\date{\today}
\begin{abstract}
We measure the ordinary and the anomalous Hall effect in a set of yttrium 
iron garnet$|$platinum (YIG$|$Pt) bilayers via magnetization orientation dependent magnetoresistance experiments. Our 
data show that the presence of the ferrimagnetic insulator YIG leads to an anomalous Hall effect like voltage in Pt, which is sensitive to both Pt thickness and temperature. Interpretation of the experimental findings in terms of the spin Hall anomalous Hall effect indicates that the imaginary part of the spin mixing conductance $G_{\mathrm{i}}$ plays a crucial role in YIG$|$Pt bilayers. In particular, our data suggest a sign change in $G_{\mathrm{i}}$ between $10\,\mathrm{K}$ and $300\,\mathrm{K}$. Additionally, we report a higher order Hall effect contribution, which appears in thin Pt films on YIG at low temperatures.
\end{abstract}
\maketitle
\begin{figure}[tb]
\includegraphics{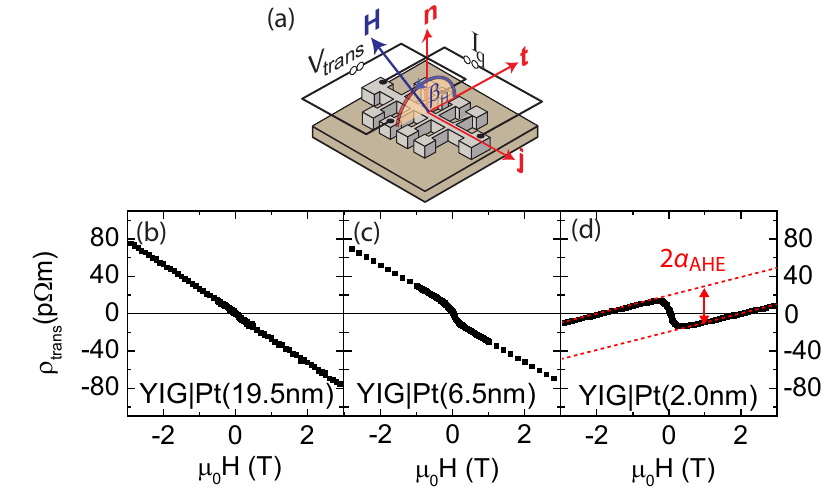}
	\caption{\textbf{(a)} Sample and measurement geometry. \textbf{(b)} - \textbf{(d)}:Transverse resistivity $\rho_{\mathrm{trans}}$ taken from FDMR measurements for YIG$|$Pt bilayers with \textbf{(b)} $t_{\mathrm{Pt}}= 19.5\,\mathrm{nm}$, \textbf{(c)} $6.5\,\mathrm{nm}$ and \textbf{(d)} $2.0\,\mathrm{nm}$, respectively. All data are taken at $300\,\mathrm{K}$. The dashed red lines in panel \textbf{(d)} indicate the extraction of $\alpha_{\mathrm{AHE}}$ from linear fits to $\rho_{\mathrm{trans}}(H)$ extrapolated to $\mu_0 H=0\,\mathrm{T}$.}
	\label{fig2}
\end{figure}

The generation, manipulation and detection of pure spin currents are fascinating challenges in \textit{spintronics}. In normal metals with large spin orbit interaction, the spin Hall effect (SHE)\cite{dyakonov1971,hirsch_spin_1999, kato_observation_2004, Valenzuela_Tinkham_2006} and its inverse (ISHE) \cite{saitoh_conversion_2006} enable the generation viz. detection of spin currents in the charge transport channel. In this context, the spin Hall angle $\theta_{\mathrm {SH}}$ and the spin diffusion length $\lambda$ are key material parameters\cite{dyakonov1971,hirsch_spin_1999}. Additionally, the spin mixing conductance $G$ was proposed as a measure for the number of spin transport channels per unit area across a normal metal (NM)$|$ferromagnet (FM) interface, in analogy to the Landauer-B\"{u}ttiker picture in ballistic charge transport \cite{Tserkovnyak_Brataas_Bauer_2002, Xia_Kelly_Bauer_Brataas_Turek_2002}. Here, $G=G_{\mathrm{r}} + \imath G_{\mathrm{i}}$ is introduced as a complex quantity \cite{Hernando_2000, Brataas2000, Stiles2002, Wang_damping_2011, Padron_damping_2011}. The real part $G_{\mathrm{r}}$ is linked to an in-plane magnetic field torque \cite{Ralph_Stiles_2008, Wang_Sun_Song_Wu_Schultheiß_Pearson_Hoffmann_2011} and accessible e.g. from spin pumping experiments \cite{saitoh_conversion_2006,Tserkovnyak_Brataas_Bauer_2002, Xia_Kelly_Bauer_Brataas_Turek_2002, Kajiwara_2010, czeschka_scaling_2011}. The imaginary part $G_{\mathrm{i}}$ is related to the spin precession and interpreted as a phase shift between the spin current in the NM and the one in the FM. $G_{\mathrm{i}}$ thus can be either positive or negative \cite{Xia_Kelly_Bauer_Brataas_Turek_2002}. As suggested recently, the spin Hall magnetoresistance (SMR)\cite{Nakayama2013, Vlietstra_APL_2013, ChenSMR2013} based on the simultaneous action of SHE and ISHE allows for quantifying $ G_{\mathrm{i}}$ from measurements of anomalous Hall-type effects (AHE) in ferromagnetic insulator$|$NM hybrids, referred to as spin Hall anomalous Hall effect (SH-AHE)\cite{ChenSMR2013}. \\
Here, we present an experimental study of ordinary and anomalous Hall-type signals observed in yttrium iron garnet ($\mathrm{Y}_{3}\mathrm{Fe}_{5}\mathrm{O}_{12}$, YIG)$|$platinum (Pt) bilayers. We discuss the film thickness and temperature dependence of the AHE signals in terms of the SH-AHE. 
While the AHE voltage observed in metallic ferromagnets usually obeys $V_H\propto M_\perp^n$ with $n=1$ and $M_\perp$ the component of the magnetization along the film normal, we observe a more complex AHE-type response with higher order terms $V_H \propto M_\perp^n$ at low temperatures in YIG$|$Pt samples with a Pt film thickness $t_{\mathrm{Pt}}\leq 5\,\mathrm{nm}$.
The higher order contributions are directly evident in our experiments, since we measure the magneto-transport response as a function of external magnetic field orientation, while conventional Hall experiments are typically performed as a function of field strength in a perpendicular field arrangement. 
For comparison, we also study thin Pt films deposited directly onto diamagnetic substrates. In these samples, we neither find a temperature dependence of the ordinary Hall-effect (OHE), nor an
AHE-type signal, not to speak of higher order AHE contributions. \\
We investigate two types of thin film structures, YIG$|$Pt bilayers and single Pt thin films on yttrium aluminum garnet ($\mathrm{Y_{3}Al_5O_{12}}$, YAG) substrates. The YIG$|$Pt bilayers are obtained by growing epitaxial YIG thin films with a thickness of $t\approx 60\,\mathrm{nm}$ on single crystalline YAG or gadolinium gallium garnet ($\mathrm{Gd}_{3}\mathrm{Ga}_{5}\mathrm{O}_{12}$,GGG) substrates using pulsed laser deposition\cite{Gepraegs_YIG2012, Alti}. In an \textit{in situ} process, we then deposit a thin polycrystalline Pt film onto the YIG via electron beam evaporation. We hereby systematically vary the Pt thickness from sample to sample in the range $1\,\mathrm{nm} \le t_\mathrm{Pt} \le 20\,\mathrm{nm}$. In this way, we obtain a series of YIG$|$Pt bilayers with fixed YIG thickness, but different Pt thicknesses. For reference, we furthermore fabricate a series of YAG$|$Pt bilayers, depositing Pt thin films with thicknesses $2\,\mathrm{nm} \le t_\mathrm{Pt} \le 16\,\mathrm{nm}$ directly onto YAG substrates. 
We employ X-ray reflectometry and X-ray diffraction to determine $t_{\mathrm{Pt}}$ and to confirm the polycrystallinity of the Pt thin films\footnote{See supplemental material at [URL will be inserted by publisher] for details.}. For electrical transport measurements, the samples are patterned into Hall bar mesa structures (width $w=80\,\mathrm{\mu m}$, contact separation $l=600\,\mathrm{\mu m}$)\cite{Meyer_Althammer_Gepraegs_2014} [c.f. Fig.\,\ref{fig2}(a)]. We current bias the Hall bars with $I_{\mathrm q}$ of up to $500\,\mu\mathrm{A}$ and measure the transverse (Hall like) voltage $V_{\mathrm{trans}}$ either as a function of the magnetic field orientation (angle dependent magnetoresistance, ADMR \cite{limmer_angle-dependent_2006, Alti}) or of the magnetic field amplitude $\mu_0 H$ (field dependent magnetoresistance, FDMR), for sample temperatures $T$ between $10\,\mathrm{K}$ and $300\,\mathrm{K}$. For all FDMR data reported below, the external magnetic field was applied perpendicular to the sample plane ($\mu_0 \mathbf H\parallel \mathbf n$, c.f. Fig.\,\ref{fig2}(a)). For the ADMR measurements, we rotate an external magnetic field of constant magnitude $1\,\mathrm{T} \leq \mu_0 H\leq 7\,\mathrm{T}$ in the plane perpendicular to the current direction $\mathbf j$ \cite{Meyer_Althammer_Gepraegs_2014}. 
Here, $\beta_{\mathrm{H}}$ is defined as the angle between the transverse direction $\mathbf{t}$ and the magnetic field $\mathbf{H}$. In all ADMR experiments, we choose $\mu_0 H$ larger than the anisotropy and the demagnetization fields of the YIG film. As a result, the YIG magnetization $\mathbf{M}$ is always saturated and oriented along $\mathbf{H}$ in good approximation. The transverse resistivity $\rho_{\mathrm{trans}}\left( \beta_{\mathrm{H}}, H \right)=V_{\mathrm{trans}}\left( \beta_{\mathrm{H}}, H \right) t_{\mathrm{Pt}}/I_{\mathrm q}$ of the Pt layer is calculated from the voltage $V_{\mathrm{trans}}\left( \beta_{\mathrm{H}} \right)$ along $\mathbf t$.\\
Figure\,\ref{fig2}(b-d) show FDMR measurements carried out at $300\,\mathrm{K}$ in YIG$|$Pt bilayers with $t_{\mathrm{Pt}}=2.0, 6.5$ and $19.5\,\mathrm{nm}$. Extracting the ordinary Hall coefficient $\alpha_{\mathrm{OHE}}=\partial \rho_{\mathrm{trans}}(H) / \partial (H)$ from the slope, we obtain $\alpha_{\mathrm{OHE}} (19.5\,\mathrm{nm})=-25.5\mathrm{p}\Omega \mathrm{m}/\mathrm{T}$ for the thickest Pt layer [see Fig.\,\ref{fig2}(b)], close to the literature value for bulk Pt \cite{Gehlhoff_1950}.
Additionally, we observe a small superimposed S-like feature around $\mu_0 H=0\,\mathrm{T}$, indicating the presence of an AHE like contribution. To quantify this contribution, we extract the full amplitude of the S-shape corresponding to an AHE like contribution $\alpha_{\mathrm{AHE}}$ from linear fits to $\mu_0H=0\,\mathrm {T}$,as indicated in Fig.\,\ref{fig2}(d). 
In the sample with $t_\mathrm{Pt}= 6.5\,\mathrm{nm}$ [Fig.\,\ref{fig2}(c)], $\alpha_{\mathrm{OHE}}$ decreases to $-23.1\mathrm{p}\Omega \mathrm{m}/\mathrm{T}$ and we find an increased $\alpha_{\mathrm{AHE}}(t_\mathrm{Pt}=6.5\,\mathrm{nm}) = (-6\pm 1)\mathrm{p}\Omega \mathrm{m}$.
For $t_{\mathrm{Pt}}=2.0\,\mathrm{nm}$ [see Fig.\,\ref{fig2}(d)], we observe $\alpha_{\mathrm{OHE}} = 7\,\mathrm{p}\Omega \mathrm{m}/\mathrm{T}$, i.e. an inversion of the sign of the OHE. 
Additionally, we find $\alpha_{\mathrm{AHE}}$ equal to $(-12\pm 1)\,\mathrm{p} \Omega \mathrm{m}$. The presence of an AHE like behavior in YIG$|$Pt samples coincides with recent reports\cite{huang_transport_2012, Vlietstra_APL_2013, Shimizu2013, Qu2014, Miao_2014, Shiomi_2014}. However, our study of $\alpha_{\mathrm{AHE}}$ as a function of platinum thickness and temperature in addition reveals a pronounced thickness dependence of $\alpha_{\mathrm{AHE}}$ for $t_{\mathrm{Pt}}\leq 10\,\mathrm{nm}$ that will be addressed below [c.f. Fig.\,\ref{fig4}(b))]. For reference, we also performed FDMR measurements on Pt thin films deposited directly onto diamagnetic YAG substrates. In these samples, we find a similar thickness dependence of the ordinary Hall-effect (OHE), but no AHE-type signal\footnotemark[1]. Thus, the sign inversion of the OHE is intimately connected to the Pt thin film regime\cite{Vlietstra_APL_2013}.\\
\begin{figure}[b]
\includegraphics{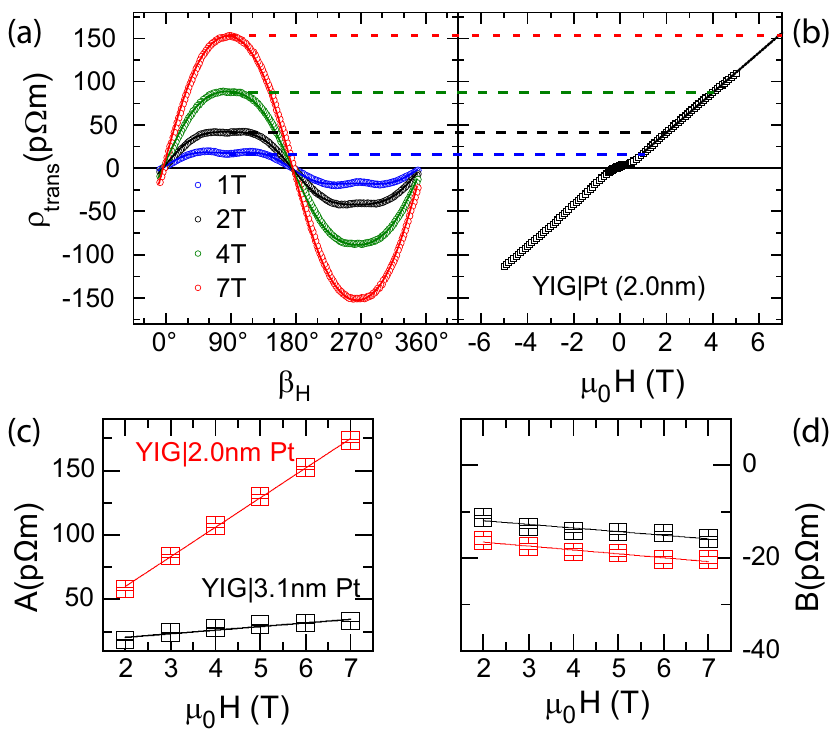}
	\caption{\textbf{(a)} ADMR and \textbf{(b)} FDMR data of a YIG$|$Pt sample with $t_\mathrm{Pt}=2.0\,\mathrm{nm}$, taken at $10\,\mathrm{K}$ for different $\mu_0H$ (open symbols). 
	The dashed horizontal lines are intended as guides to the eye, to show that the $\rho_{\mathrm{trans}}$ values inferred from FDMR and ADMR are consistent for identical magnetic field configurations.The fits of Eq.\,(\ref{eq1}) to the data are shown as solid lines. \textbf{(c)} and \textbf{(d)} show the fit parameters $A$ and $B$ obtained from Eq.\,(\ref{eq1}) for YIG$|$Pt($3.1\,\mathrm{nm}$) (black) and YIG$|$Pt($2.0\,\mathrm{nm}$) (red) at $T=10\,\mathrm{K}$. Linear fits to the magnetic field dependence of $A$ and $B$ are shown as solid lines.}
	\label{fig3}
\end{figure}
Complementary to the FDMR experiments, we further investigate $\rho_{\mathrm{trans}}$ as a function of the magnetic field orientation (ADMR). In Fig.\,\ref{fig3}(a) we show ADMR data for a YIG$|$Pt($2.0\,\mathrm{nm}$) hybrid recorded at $10\,\mathrm{K}$. In ADMR experiments, the OHE is expected to depend only on the component $H_{\perp}=H \sin(\beta_{\mathrm{H}})$, i.e., $\rho(\beta_\mathrm{H})\propto \sin(\beta_\mathrm{H})$. 
However, our experimental data reveals additional higher than linear order contributions of the form $V_\mathrm{trans} \propto M_\perp^n$, with $\rho_{\mathrm{trans}} \propto A\sin(\beta_{\mathrm{H}})+B\sin^3(\beta_{\mathrm{H}})+ \cdots$. A fast Fourier transformation\footnotemark[1]
of the ADMR data suggests the presence of $\sin^n(\beta_{\mathrm{H}})$ contributions up to at least $n=5$\footnotemark[1]. However, a quantitative determination of corresponding higher order coefficients is difficult, since the amplitudes of the contributions for $n\geq 5$ are below our experimental resolution of $1\,\mathrm{p}\Omega\mathrm{m}$.
A behavior similar to that shown in Fig.\,\ref{fig3}(a) is found in all YIG$|$Pt samples with $t_{\mathrm{Pt}}\leq 5\,\mathrm{nm}$, but not in plain Pt films on YAG\footnotemark[1]. To allow for simple analysis, we use
\begin{equation}
\rho_{\mathrm{trans}}=A \sin(\beta_{\mathrm{H}})+B \sin^3(\beta_{\mathrm{H}})
\label{eq1}
\end{equation}
 in the following. Fits of the ADMR curves measured at different field magnitudes according to Eq.\,(\ref{eq1}) are shown as solid lines in Fig.\,\ref{fig3}(a). The magnetic field dependence of the fit parameters $A$ and $B$ is shown in Figs.\,\ref{fig3}(c),(d) for two samples with $t_{\mathrm{Pt}}=3.1\,\mathrm{nm}$\footnotemark[1] and $t_{\mathrm{Pt}}=2.0\,\mathrm{nm}$.
We disentangle magnetic field dependent (OHE like) and "field independent" (AHE like) contributions to $A$ by fitting the data to $A (\mu_0H)=A_{\mathrm{OHE}} \mu_0 H + A_{\mathrm{AHE}}$. As evident from Fig.\,\ref{fig3}, the $\alpha_{\mathrm{OHE}}$ and $\alpha_{\mathrm{AHE}}$ values derived from FDMR and ADMR measurements are quantitatively consistent.\\
The $A_{\mathrm{OHE}}$ as a function of $t_{\mathrm{Pt}}$ is shown in Fig.\,\ref{fig4}(a). Obviously, $A_{\mathrm{OHE}}$ deviates from the bulk OHE literature value \cite{Gehlhoff_1950} in YIG$|$Pt bilayers with $t_{\mathrm{Pt}}\leq 10\,\mathrm{nm}$ and also exhibits a temperature and thickness-dependent sign change for small $t_{\mathrm{Pt}}$. A thickness-dependent behavior of the OHE without sign change has also been reported in Ref.\,\onlinecite{Vlietstra_APL_2013}. However, these authors found an increase of the OHE coefficient in the thin film regime, which could be due to the formation of a thin, non-conductive ``dead'' Pt layer at the interface as, e.g., reported for Ni$|$Pt\cite{Shin1998}. In contrast, we attribute the thickness dependence of the OHE in our samples solely to a modification of the Pt properties in the thin film regime. Further experiments will be required in the future to clarify the origin of the temperature dependence of the OHE in YIG$|$Pt hybrids.\\
The anomalous Hall coefficient $A_{\mathrm{AHE}}$, present only in YIG$|$Pt hybrids, i.e., when a magnetic insulator is adjacent to the NM, is depicted in Fig.\,\ref{fig4}(b). We observe a strong dependence of $A_{\mathrm{AHE}}$ on $t_{\mathrm{Pt}}$ similar to the thickness dependent magnetoresistance obtained from longitudinal transport measurements reported earlier\cite {Alti}, but with a sign change in $A_{\mathrm{AHE}}$ between $100\,\mathrm{K}$ and $10\,\mathrm{K}$. This observation agrees with recent reports of $A_\mathrm{AHE} = 54\,\mathrm p \Omega \mathrm m$ for YIG$|$Pt($1.8\,\mathrm{nm}$) \cite{Shiomi_2014} and $A_\mathrm{AHE} = 6\,\mathrm p \Omega \mathrm m$ for YIG$|$Pt($3\,\mathrm{nm}$) \cite{Miao_2014}, both taken at $10\,\mathrm K$. 
Our study suggests a maximum in $A_{\mathrm{AHE}}$ around $t_{\mathrm{Pt}}=3\,\mathrm{nm}$, compatible with a complete disappearance of $A_{\mathrm{AHE}}$ for $t_{\mathrm{Pt}}\rightarrow 0$.
 This observation however is at odds with the attribution of the AHE in YIG$|$Pt to a proximity MR as postulated in Ref.\,\onlinecite{Miao_2014}. In this case one would expect a monotonous increase of the AHE signal with decreasing Pt layer  thickness, and eventually a saturation when the entire nonmagnetic layer is spin polarized. The absence of a proximity MR in our Hall data is consistent with XMCD data on similar YIG$|$Pt samples \cite{Gepraegs_YIG2012} as well as other ferromagnetic insulator$|$NM hybrids\cite{Satapathy2012}. However, we want to point out that a magnetic proximity effect has been reported in some YIG$|$Pt samples\cite{Lu_XMCD_2013, Lu_MR_2013}.\\
We now model our experimental findings in terms of the SH-AHE theory \cite{ChenSMR2013} 
\begin{equation}
\rho_{\mathrm{trans}}=-\frac{2\lambda^2\theta_{\mathrm{SH}}^2}{t_{\mathrm{Pt}}} \frac {G_{\mathrm i}\tanh^2\left( \frac{t_{\mathrm{Pt}}}{2\lambda}\right)}{(\sigma + 2\lambda G_{\mathrm{r}}\coth \left(\frac{t_{\mathrm{Pt}}}{\lambda}\right))^2} m_n,
\label{eq2}
\end{equation}
where $\sigma=\rho^{-1}$ is the electric conductivity of the Pt layer and $m_n$ the unit vector of the projection of the
magnetization orientation $\mathbf{m}$ onto the direction $\mathbf{n}$ (c.f. Fig.\,\ref{fig2}). To fit the nonlinear behavior of 
$A_{\mathrm{AHE}}(t_{\mathrm{Pt}})$, we combine this expression with the thickness dependence of the sheet resistivity for thin Pt films \cite{Fischer1980} as discussed in Ref.\,\onlinecite{Meyer_Althammer_Gepraegs_2014}. We use the parameters $\lambda=1.5\,\mathrm{nm}$, $G_{\mathrm{r}}=4\times10^{14}\Omega^{-1}\mathrm{m}^{-2}$, $\theta_{\mathrm{SH}} (300\,\mathrm{K})=0.11$ and $\theta_{\mathrm{SH}} (10\,\mathrm{K})=0.07$ obtained from longitudinal SMR measurements on similar YIG$|$Pt bilayers \cite{Meyer_Althammer_Gepraegs_2014}. As obvious from the solid lines in Fig.\,\ref{fig4}(b), Eq.\,(\ref{eq2}) reproduces our thickness dependent AHE data upon using $G_{\mathrm{i}}=1\times 10^{13}\,\Omega^{-1}\mathrm{m}^{-2}$ for $300\,\mathrm{K}$ and $G_{\mathrm{i}}=-3\times 10^{13}\,\Omega^{-1}\mathrm{m}^{-2}$ for $10\,\mathrm{K}$. For $300\,\mathrm{K}$, the value for $G_{\mathrm{i}}$ nicely coincides with earlier reports\cite{Alti} as well as theoretical calculations\cite{jia_spin_2011}. In the SH-AHE model, the only parameter allowing to account for the sign change in $\rho_{\mathrm{trans}}$ as a function of temperature is $G_{\mathrm{i}}$. In this picture, our AHE data thus indicate a sign change in $G_{\mathrm{i}}$ between $300\,\mathrm{K}$ and $10\,\mathrm{K}$.\\
\begin{figure}[t]
\includegraphics{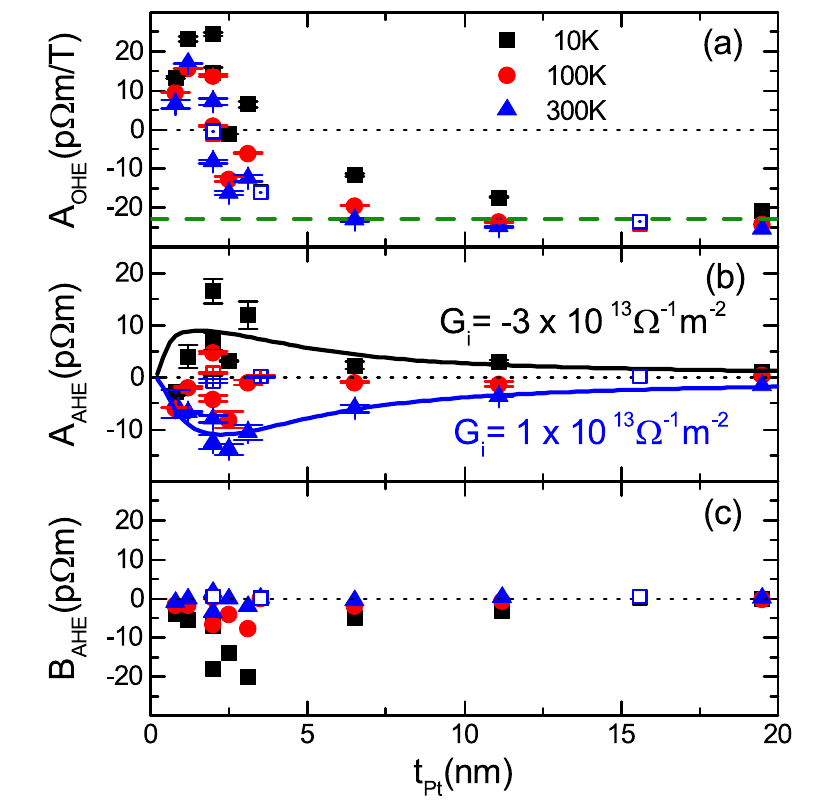}
	\caption{\textbf{(a)-(c)} Field dependent (OHE-like) and field independent (AHE-like) Hall coefficients $A$ and $B$ proportional to $\sin{(\beta_{\mathrm{H}})}$ and $\sin^3{(\beta_{\mathrm{H}})}$, respectively, plotted versus the Pt thickness for $T=300\,\mathrm K$ (blue), $T=100\,\mathrm K$ (red) and $T=10\,\mathrm K$ (black). The data is obtained from ADMR measurements for YAG$|$Pt (open symbols) and YAG$|$YIG$|$Pt (full symbols). $A_{\mathrm{OHE}}$ depicted in \textbf{(a)} describes the conventional Hall effect, the olive dashed line corresponds to the literature value for bulk Pt\cite{Gehlhoff_1950}. \textbf{(b) }Thickness dependence of $A_{\mathrm{AHE}}$. The solid lines show fits to the SH-AHE theory using $G_i=1 \times 10^{13}\Omega^{-1}\mathrm{m}^{-2}$ for $T=300\,\mathrm K$ (blue) and $G_i=-3 \times 10^{13}\Omega^{-1}\mathrm{m}^{-2}$ for $T=10\,\mathrm K$ (black). Panel \textbf{(c)} shows the thickness dependence of the field independent coefficient $B_{\mathrm{AHE}}$ of the $\sin^3{(\beta_{\mathrm{H}})}$ term.}
	\label{fig4}
\end{figure}
We finally address the thickness and temperature dependence of the $\sin^3(\beta_{\mathrm{H}})$ contribution parametrized by $B=B_{\mathrm{AHE}}+ B_{\mathrm{OHE}} \mu_0 H$, that cannot straightforwardly be explained in a conventional Hall scenario. As evident from the linear fits in Fig.\,\ref{fig3}(c), $B$ is nearly field independent. A slight field dependence $B_{\mathrm{OHE}}\leq 1\,\mathrm{p}\Omega \mathrm{m}/\mathrm{T}$ might arise due to fitting errors caused by neglected higher order terms $(n\geq 5)$. Therefore, we focus our discussion on the field independent part $B_{\mathrm{AHE}}$ in the following. 
$B_{\mathrm{AHE}}$ exhibits a strong temperature and thickness dependence as shown in Fig.\,\ref{fig4}(c), suggesting a close link to $A_{\mathrm{AHE}}$ and therefore the SH-AHE. However, we do not observe a temperature-dependent sign change in $B_{\mathrm{AHE}}$. Expanding the SMR theory\cite{ChenSMR2013} to include higher order contributions of the magnetization directions $m_i$ ($i=j, t, n$) in analogy to the procedure established for the AMR of metallic ferromagnets\cite{limmer_angle-dependent_2006, limmer_advanced_2008}, $\sin^3(\beta_\mathrm H)$ terms appear in $\rho_\mathrm{trans}$, but with an amplitude proportional to $\theta_{\mathrm{SH}}^4$. Assuming $\theta_{\mathrm{SH}}(\mathrm{Pt})\approx 0.1$, this would lead to $B_{\mathrm{AHE}}/A_{\mathrm{AHE}}\approx 0.01$, which disagrees with our experimental finding $B_{\mathrm{AHE}}/A_{\mathrm{AHE}}\geq 0.2$. 
Additionally, we study the influence of the longitudinal resistivity on $\rho_{\mathrm{AHE}}$. For metallic ferromagnets, one usually considers $\rho_{\mathrm{AHE}}\propto M(H) \rho_{\mathrm{long}}^{\alpha}$ with $1 \leq \alpha \leq 2$ \cite{Karplus_1954, Berger_1970}. Applying this approach to $V_\mathrm{trans}$ of the YIG$|$Pt samples discussed here is not possible: Since the longitudinal resistance is modulated by the SMR with $\rho_1/\rho_0\leq 10^{-3}$ \cite{Alti}, $\rho_{\mathrm{AHE}}\propto \rho^{\alpha}$ would imply $B_{\mathrm{AHE}}/A_{\mathrm{AHE}}\leq 10^{-3}$. This is in contrast to our experimental findings. Thus, a dependence of the form $\rho_\mathrm{AHE} \propto \rho_{\mathrm{long}}^{\alpha}$ cannot account for our experimental observations.
Finally, a static magnetic proximity effect\cite{huang_transport_2012, Lu_XMCD_2013, Lu_MR_2013} also cannot explain $B_{\mathrm{AHE}}$, since the thickness dependence of $B_{\mathrm{AHE}}$ shown in Fig.\,\ref{fig4} (c) clearly indicates a decrease for $t_{\mathrm{Pt}}\leq 2.5\,\mathrm{nm}$. Consequently, within our present knowledge, neither a spin current related phenomenon (SMR, SH-AHE), nor a proximity based effect can explain the origin or the magnitude of this anisotropic higher order anomalous Hall effect.  We also would like to point out that the higher order $\sin^3(\beta_{\mathrm{H}})$ term can be resolved only in ADMR measurements. In conventional FDMR experiments, such higher order contributions cannot be discerned.\\
In summary, we have investigated the anomalous Hall effect in YIG$|$Pt heterostructures for different Pt thicknesses, comparing magnetization orientation dependent (ADMR) and magnetic field magnitude dependent (FDMR) measurements at temperatures between $10\,\mathrm{K}$ and $300\,\mathrm{K}$. 
In Pt thin films on diamagnetic (YAG) substrates, we observe a Pt thickness dependent ordinary Hall effect (OHE) only. However, in YIG$|$Pt bilayers, an AHE like signal is present in addition. The AHE effect changes sign as a function of temperature and can be modeled using a spin Hall magnetoresistance-type formalism for the transverse transport coefficient. However, we need to assume a sign change in the imaginary part of the spin mixing interface conductance to describe the sign change in the anomalous Hall signal observed experimentally. Finally, we identify contributions proportional to $\sin^3(\beta_{\mathrm{H}})$ and higher orders in the ADMR data for YIG$|$Pt. The physical mechanism responsible for this behavior could not be clarified within this work and will be subject of further investigations. The observation of higher order contributions to the AHE in angle dependent magnetotransport measurements confirms the usefulness of magnetization orientation dependent experiments. Clearly, magnetotransport measurements as a function of the magnetic field magnitude only, i.e. for a single magnetic field orientation (perpendicular field), as usually performed to study Hall effects, are not sufficient to access all transverse transport features. \\[0.5cm]
We thank T. Brenninger for technical support and M. Schreier for fruitful discussions.
Financial support by the Deutsche Forschungsgemeinschaft via SPP 1538 (project no. GO 944/4) is gratefully acknowledged.
\bibliography{BiblioAbbrev}
\pagebreak
\begin{center}
\textbf{\large Supplemental Materials: Anomalous Hall effect in YIG$|$Pt bilayers}
\end{center}
\section {Reference measurements in YAG$|$Pt bilayers}
\begin{figure}[h]
	\includegraphics{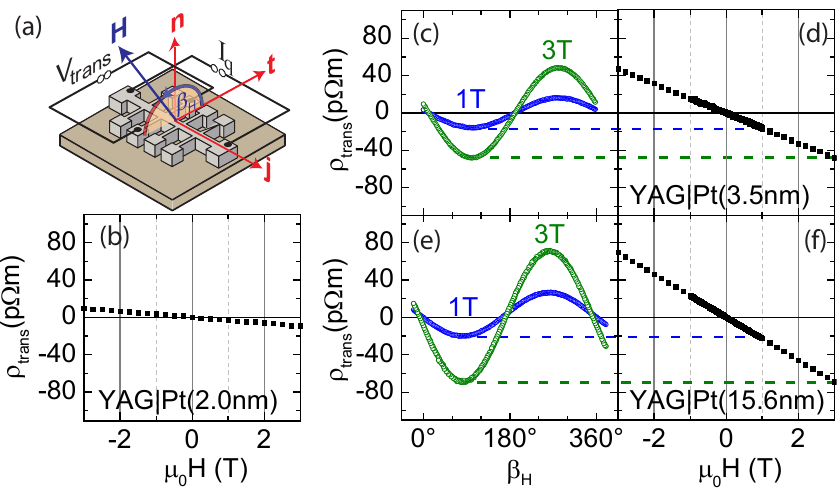}
	\caption{\textbf{(a)} Sample and measurement geometry. \textbf{(b)} Transverse resistivity $\rho_{\mathrm{trans}}$ taken from a FDMR measurement for YAG$|$Pt ($2.0\,\mathrm{nm}$). \textbf{(c)} $\rho_{\mathrm{trans}}$ as a function of $\beta_{\mathrm{H}}$ for YAG$|$Pt ($3.5\,\mathrm{nm}$). \textbf{(d)} Corresponding FDMR data for $\beta_{\mathrm{H}}=90^\circ$.
\textbf{(e, f)}: ADMR and FDMR measurements for $t_{\mathrm{Pt}}=15.6\,\mathrm{nm}$ on YAG. 
The colored, horizontal, dashed lines in panels (c,d) and (e,f) are intended as guides to the eye, to show that the $\rho_{\mathrm{trans}}$ values inferred from FDMR and ADMR are consistent for identical magnetic field configurations. All data taken at $300\,\mathrm{K}$.}
	\label{S1}
\end{figure}

Here, we discuss the reference samples consisting of plain Pt thin films on single-crystalline diamagnetic Yttrium Aluminum Garnet (YAG). Figures \,S\,\ref{S1}(b),(d),(f) show the characteristic linear behavior for $\rho_{\mathrm{trans}} (H)\propto H$, i.e., an ordinary Hall effect, without any AHE contribution. Extracting the ordinary Hall coefficient $\alpha_{\mathrm{OHE}}=\partial \rho_{\mathrm{trans}}( H) / \partial ( H)$ from the slope, we obtain $\alpha_{\mathrm{OHE}}(t_{\mathrm{Pt}}=2.0\,\mathrm{nm})=-3.1\,\mathrm{p}\Omega \mathrm{m}/\mathrm{T}$, $\alpha_{\mathrm{OHE}}(t_{\mathrm{Pt}}=3.5\,\mathrm{nm})=-15.9\,\mathrm{p}\Omega \mathrm{m}/\mathrm{T}$, and $\alpha_{\mathrm{OHE}}(t_{\mathrm{Pt}}=15.6\,\mathrm{nm})=-23.1\,\mathrm{p}\Omega \mathrm{m}/\mathrm{T}$ with a systematic error of $\Delta \alpha_\mathrm{OHE}=0.1\,\mathrm{p}\Omega \mathrm{m}/\mathrm{T}$.
While $\alpha_\mathrm{OHE}$ of the thickest Pt film with $t_\mathrm{Pt}=15.6\,\mathrm{nm}$ is consistent with the literature value $\alpha_{\mathrm{OHE}}=-24.4\,\mathrm{p}\Omega \mathrm{m}/\mathrm{T}$ \cite{Gehlhoff_1950}, we find significantly smaller OHE coefficients for the $3.5\,\mathrm{nm}$ and the $2.0\,\mathrm{nm}$ thick Pt film.
This behavior in the thin film regime $\left (t_{\mathrm{Pt}}\leq 10\,\mathrm{nm} \right)$ agrees with earlier reports\cite{Panchenko1969}.
 Measurements of $\alpha_{\mathrm{OHE}}(T)$ show a $T$ independent $\alpha_{\mathrm{OHE}}$ and the absence of any AHE like contribution. 
Complementary to the FDMR experiments, we investigate $\rho_{\mathrm{trans}}$ as a function of the magnetic field orientation (ADMR). As evident from Fig.\,S\,\ref{S1}(c) and (e), we obtain that the OHE depends only on the component $H_{\perp}=H \sin(\beta_{\mathrm{H}})$, i.e., $\rho_\mathrm{trans}(\beta_\mathrm{H})\propto \sin(\beta_\mathrm{H})$.
\section {Magnetization Orientation and Field Magnitude Dependent Measurements for YIG$|$Pt $\mathbf{(3.1\,nm)}$}
\begin{figure}[h!]
	\includegraphics[width=\columnwidth]{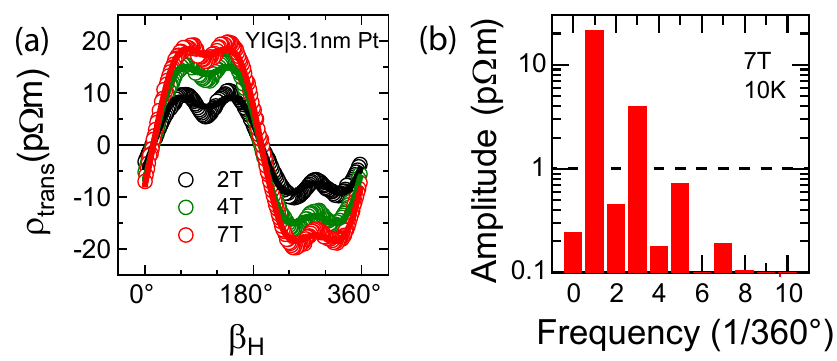}
	\caption{\textbf{(a)}Transverse resistivity $\rho_{\mathrm{trans}}$ as a function of $\beta_{\mathrm{H}}$ for a YIG$|$Pt bilayer with $t_{\mathrm{Pt}}=3.1\,\mathrm{nm}$, taken at $10\,\mathrm{K}$.\textbf{(b)}Fast Fourier transform (FFT) of the ADMR data taken at $T=10\,\mathrm K$ with $\mu_0H=7\,\mathrm{T}$ for the YIG$|$Pt($3.1\,\mathrm{nm}$) sample shown in \textbf{(a)}. The dashed line indicates the experimental noise level of $1\,\mathrm p \Omega\mathrm m$.} 
	\label{S2}
\end{figure}
In Fig.\,S\,\ref{S2}(a) we show a set of ADMR data for a YIG$|$Pt ($3.1\,\mathrm{nm}$) sample taken at $10\,\mathrm{K}$ as a reference to Fig.\,2 in the main text. This additional data substantiates the reproducibility of the observation of higher order contributions to $\rho_{\mathrm{trans}}$ up to at least $n=3$ in a set of ADMR measurements [see Fig.\,S\,\ref{S2}]. 
Please note that the data shown in Fig.\,S\,\ref{S2}(a) is taken at $10\,\mathrm{K}$, while the FDMR measurements performed on similar samples shown in Fig.\,1 were taken at $300\,\mathrm{K}$ and thus have a different OHE and AHE behavior. 
In particular, for $T=10\,\mathrm K$, we observe an almost vanishing OHE signal in this sample, $\alpha_{\mathrm{OHE}}=6\,\mathrm{p}\Omega \mathrm{m}/\mathrm{T}$ and therefor the $\sin^3(\beta_{\mathrm{H}})$ contribution becomes prominent even for the $7\,\mathrm{T}$ data, which otherwise would be overwhelmed by the $\sin(\beta_{\mathrm{H}})$ characteristic of the OHE.
The fitting parameters $A$ and $B$ obtained from fits of Eq.(1) to the ADMR data shown in S\,\ref{S2}(a) for YIG$|$Pt ($3.1\,\mathrm{nm}$) are represented by the black data points in Fig.\,2(c) and (d) in the main article. For a full picture of the temperature dependence of OHE and AHE contributions to the parameters $A$ and $B$, we refer to Fig.\,3 in the main text. 
\section {Fast Fourier Transform}
As shown in Fig.\,S\,\ref{S2}(a) and in Fig.\,2(a) in the main text, our magnetization orientation dependent measurements on YIG$|$Pt bilayers reveal additional higher order contributions to $\rho_{\mathrm{trans}}$, such that we can formulate $\rho_{\mathrm{trans}} \propto A\sin(\beta_{\mathrm{H}}) + B\sin^3(\beta_{\mathrm{H}})+\cdots$. To specify the particular contributions, we perform fast Fourier transformations (FFT) of the ADMR data as exemplarily shown in Fig.\,S\,\ref{S2}(b) for YIG$|$Pt($3.1\,\mathrm {nm}$) taken at $10\,\mathrm K$ [see Fig.\,S\,\ref{S2}(a)]. For the FFT, we use a rectangular window with amplitude correction. The amplitude spectrum of the FFT for this set of data reveals the presence of $\sin(n\beta_\mathrm{H})$ contributions up to at least $n=5$. Possibly occurring higher order contributions could not be quantified, since the amplitude for the $n=5$ contribution is already comparable to our experimental resolution of $1\,\mathrm p \Omega \mathrm m$.\\
Please note that the FFT results depicted in S\,\ref{S2}(b) are not sign-sensitive and can not straightforwardly be compared to results for $A$ and $B$ obtained from fits using Eq.\,(2). The FFT algorithm specifies frequency components proportional to $\sin(n\beta_\mathrm{H})$, while our approximation in Eq.\,(2) is a power series proportional to $\sin^n(\beta_\mathrm{H})$. However, both expressions represent the same phenomenology and can be transformed into the respective other by fundamental algebra.
\newpage
\section {Table of Samples}
A detailed information on the film thicknesses for both types of thin film structures used in our study is listed in Tab.\,S\,\ref{tab:samples}. The parameter $h$ represents the surface roughness of Pt obtained from high-resolution X-ray reflectometry. For YIG$|$Pt bilayers, we determine an averaged surface roughness of $h =\left(0.7 \pm 0.2\right)\,\mathrm{nm}$, while for plain Pt on diamagnetic substrate, we obtain a slightly lower value of $h =\left(0.5 \pm 0.1\right)\,\mathrm{nm}$.
However, within the estimated errors, the interface roughnesses of both types of samples are comparable and thus we expect no influence of the surface roughnesses on our OHE and AHE data.
\begin{table}[h!]
\begin{center}
		\begin{tabular}{c c c c }
  \hline
	\hline
  substrate & $t_\mathrm{YIG} (\mathrm{nm})$ & $t_\mathrm{Pt} (\mathrm{nm})$&$h (\mathrm{nm})$\\
			\hline 
  YAG & 34 & 0.8 & 0.7 \\ 
	YAG & 56 & 3.1 & 1.0 \\ 
	YAG & 38 & 1.2 & 0.9 \\
	YAG & 63 & 6.5 & 0.9 \\ 
  YAG & 57 & 2.0 & 0.8 \\
	GGG & 61 & 11.1 & 0.6 \\ 
  YAG & 49 & 2.0 & 0.6 \\ 
	YAG & 61 & 19.5 & 1.0 \\ 
  YAG & 58 & 2.5 & 1.1 \\ 
	\hline
	\hline
  YAG & 0 & 2.0 & 0.4 \\ 
	YAG & 0 & 15.6 & 0.6 \\ 
  YAG & 0 & 3.5 & 0.5 \\ 
	\hline
		\end{tabular}
	\caption{Substrate material, YIG thickness $t_\mathrm{YIG}$, platinum thickness $t_\mathrm{Pt}$ and platinum roughness $h$ for all samples investigated in this work.}
	\label{tab:samples}
\end{center}
\end{table}
\bibliography{BiblioAbbrev}
\newpage
\end{document}